\documentclass[prd,twocolumn]{revtex4}
\usepackage{graphicx, epsfig}
\usepackage{color}
\usepackage{mathrsfs}
\usepackage{amsmath}

\newcommand{\be}{\begin{equation}}
\newcommand{\ee}{\end{equation}}
\newcommand{\bea}{\begin{eqnarray}}
\newcommand{\eea}{\end{eqnarray}}

\newcommand{\gapp}{\mathrel{\raise.3ex\hbox{$>$}\mkern-14mu
              \lower0.6ex\hbox{$\sim$}}}
\newcommand{\lapp}{\mathrel{\raise.3ex\hbox{$<$}\mkern-14mu
              \lower0.6ex\hbox{$\sim$}}}

\begin{document}

\title{Schrodinger formalism, black hole horizons and singularity behavior}
\author{John E. Wang}
\affiliation{Department of Physics, Niagara University, Niagara, NY 14109-2044}
\affiliation{HEPCOS, Department of Physics,
SUNY at Buffalo, Buffalo, NY 14260-1500}
\author{Eric Greenwood}
\author{Dejan Stojkovic}
\affiliation{HEPCOS, Department of Physics,
SUNY at Buffalo, Buffalo, NY 14260-1500}

\begin{abstract}
The Gauss-Codazzi method is used to discuss the gravitational collapse of a charged Reisner-Nordstr\"om domain wall.  We solve the classical equations of motion of a thin charged shell moving under the influence of its own gravitational field and show that a form of cosmic censorship applies.  If the charge of the collapsing shell is greater than its mass, then the collapse does not form a black hole. Instead, after reaching some minimal radius, the shell bounces back.  The Schrodinger canonical formalism is used to quantize the motion of the charged shell.  The limits near the horizon and near the singularity are explored.  Near the horizon, the Schrodinger equation describing evolution of the collapsing shell takes the form of the massive wave equation with a position dependent mass. The outgoing and incoming modes of the solution are related by the Bogolubov transformation which precisely gives the Hawking temperature.  Near the classical singularity, the Schrodinger equation becomes non-local, but the wave function describing the system is non-singular.  This indicates that while quantum effects may be able to remove the classical singularity, it may also introduce some new effects.
\end{abstract}

\maketitle

\section{Introduction}

Gauss-Codazzi equations are fundamental equations in the theory of surfaces embedded in a higher dimensional space. They provide a very powerful tool of studying problems in general relativity.
Most of the work in the existing literature focused on problems with sources containing only mass distributions. The next natural step is to generalize the Gauss-Codazzi method so that the stress-energy sources with both mass and charge can be included. First, we will setup the formalism and derive the equations of motion for a charged two dimensional surface. The conserved quantities will follow from these equations of motions.  A charged shell of matter (represented by a domain wall) moves under the influence of its own gravitational and electromagnetic field. If the mass of the shell is greater than its charge, the collapse will end by the formation of a black hole.  When the charge is greater than the mass parameter, the collapse will not yield a black hole, in agreement with the cosmic censorship conjecture. In this case, the solution becomes oscillatory. The shell will collapse to some minimal radius at which the electromagnetic repulsion will overcome the gravitational attraction and cause the bounce. From that moment on, the shell will be expanding until it reaches some maximal radius at which the gravitational force will again dominate, and the new collapsing cycle will start.

We then quantize the motion of the charged shell in the context of the canonical formalism. Two of the most important regimes will be the limits near the horizon and near the singularity.  Near the horizon, the Schrodinger equation describing evolution of the collapsing shell takes the form of the massive wave equation in a Minkowski background with a position dependent mass.  Incoming and outgoing modes are defined and related by the Bogolubov transformation. Despite the fact that the outgoing state is a pure state, it has a Boltzmann distribution at the Hawking temperature. In the absence of the matter fields propagating in the background of the collapsing shell, this is an intriguing result, indicating perhaps that the collapsing shell loses its mass in the form of emitted gravitons or pair production of shells.  Near the classical singularity, the Schrodinger equation becomes non-local, but the wave function describing the system is non-singular.  We find that locality is recovered in the limit of a very large domain wall tension (i.e. mass of the collapsing shell) but negligible gravitational interaction.  Including gravitational interactions, the system is manifestly non-local.

\section{The Gauss-Codazzi Formalism}
\subsection{The Equations}
Here we setup the Einstein's equations in the presence of stress-energy sources with both mass and charge confined to three-dimensional time-like hypersurfaces.
We follow the technique developed in Ref.~\cite{Ipser:1983db}. Let $S$ denote a three-dimensional time-like hypersurface containing stress-energy and let $\zeta^a$ be its unit space-like normal ($\zeta_a\zeta^a=1$). The three-metric intrinsic to the hypersurface $S$ is
\be
  h_{ab}=g_{ag}-\zeta_a\zeta_b
\ee
where $g_{ab}$ is the four-metric of the space-time. Let $\nabla_a$ denote the covariant derivative associated with $g_{ab}$ and let
\be
  D_a=h_a{}^b\nabla_b
\ee
where $D_a$ is a projection into the hypersurface $S$ of the covariant derivative $\nabla_a$ of space-time, and $h_{ac}$ is the induced metric on the hypersurface $S$.
The extrinsic curvature of $S$, denoted by $\pi_{ab}$, is defind by
\be
  \pi_{ab}\equiv D_a\zeta_b=\pi_{ba}.
\ee
The contracted forms of the first and second Gauss-Codazzi equations are then given by
\bea
  ^3R+\pi_{ab}\pi^{ab}-\pi^2&=&-2G_{ab}\zeta^a\zeta^b\label{Gauss}\\
  h_{ab}D_c\pi^{ab}-D_a\pi&=&G_{bc}H^b{}_a\zeta^c\label{Codazzi}.
\eea
Here $^3R$ is the Ricci scalar curvature of the three-geometry $h_{ab}$ of $S$, $\pi$ is the trace of the extrinsic curvature, and $G_a{}^b$ is the Einstein tensor in four-dimensional space-time.

The stress-energy tensor $T_{ab}$ of four-dimensional space-time has a $\delta$-function singularity on $S$ for both the mass and the charge. This in turn implies that the extrinsic curvature has a jump discontinuity across $S$, since the extrinsic curvature is analogous to the gradient of the Newtonian gravitational potential. Therefore we can introduce
\be
  \gamma_{ab}\equiv\pi_{+ab}-\pi_{-ab}
\ee
and
\be
  S_{ab}\equiv\int dl \ T_{ab},
\ee
where $l$ is the proper distance through $S$ in the direction of the normal $\zeta^a$, and where the subscripts $\pm$ refer to values just off the surface on the side determined by the direction of $\pm\zeta^a$. Using the Einstein and the Gauss-Codazzi equations, one then has
\be
  S_{ab}=-\frac{1}{8\pi G_N}\left(\gamma_{ab}-h_{ab}\gamma_c{}^c\right).
  \label{action}
\ee
We can also introduce the ``average" extrinsic curvature
\be
  \tilde{\pi}_{ab}=\frac{1}{2}\left(\pi_{+ab}+\pi_{-ab}\right).
\ee
Then, using Eq. (\ref{action}), by adding and subtracting Eq. (\ref{Gauss}) and Eq. (\ref{Codazzi}) on opposite sides of $S$ we get
\begin{align}
  h_{ac}D_bS^{cb}=&0,\\
  h_{ac}D_b\tilde{\pi}^{cb}-D_a\tilde{\pi}=&0,\\
  %\tilde{\pi}_{ab}S^{ab}=&0\\
  ^3R+\left(\tilde{\pi}_{ab}\tilde{\pi}^{ab}-\tilde{\pi}^2\right)=&-16\pi^2G^2_N\left[S_{ab}S^{ab}-\frac{1}{2}(S_a{}^a)^2\right].
\end{align}
These from a complete set of equations to solve Einstein's equations in the presence of a thin wall.

%\subsection{The Surface Stress-Energy Tensor}
%
%Here we restrict ourselves to sources for which the stress energy tensor is %given by, see Refs. \cite{Bekenstein}
%\be
%  S^{ab}=\sigma %u^au^b-\tau\left(h^{ab}+u^au^b\right)+\frac{1}{4}\pi\left[F^{ac}F^b{}_c-\frac{1}{4}h^{ab}F^{cd}F_{cd}\right]
%  \label{Stress-Energy}
%\ee
%which is the sum of a part for the perfect fluid and a part for the %electromagnetic field. In Eq. (\ref{Stress-Energy}) $u^a$ is the four-velocity %of any observer whose world line lies within $S$ and who sees no energy flux in %his local frame, and where $\sigma$ is the energy per unit area and $\tau$ is %the tension measured by the observer. For a dust wall it is well known that %$\tau=0$, while for a domain wall $\tau=\sigma$. For a domain wall Eq. %(\ref{Stress-Energy}) reduces to
%\be
%  S^{ab}=-\sigma %h^{ab}+\frac{1}{4}\pi\left[F^{ac}F^b{}_c-\frac{1}{4}h^{ab}F^{cd}F_{cd}\right].
%\ee

\subsection{Attractive Energy}

Here we derive equations for an observer who is hovering just above the surface $S$ on either side. Let the vector field $u^a$ be extended off $S$ in a smooth fashion. The acceleration
\bea
  u^a\nabla_au^b&=&(h^b{}_c+\zeta^b\zeta_c)u^a\nabla_au^c\nonumber\\
     &=&h^b{}_cu^a\nabla_au^c-\zeta^bu^au^c\pi_{ab}
\eea
has a jump discontinuity across $S$ since the extrinsic curvature has such a discontinuity. The perpendicular components of the accelerations of observers hovering just off $S$ on either side satisfy
\begin{align}
  \zeta_bu^a\nabla_au^b\Big{|}_++\zeta_bu^a\nabla_au^b\Big{|}_-=&-2u^au^b\tilde{\pi}_{ab} -2\frac{1}{\sigma}S^{ab}\tilde{\pi}_{ab}\nonumber\\
     =&-2\frac{\tau}{\sigma}(h^{ab}+u^au^b)\tilde{\pi}_{ab} \nonumber \\ &-2\frac{1}{\sigma}S^{ab}\tilde{\pi}_{ab}
     \label{perp1}
\end{align}
and
\bea
   \zeta_bu^a\nabla_au^b\Big{|}_+-\zeta_bu^a\nabla_au^b\Big{|}_-&=&-u^au^b\gamma_{ab}\nonumber\\
     &=&4\pi G_n(\sigma-2\tau).
     \label{perp2}
\eea

\section{Model}

We consider a spherical domain wall with the constant tension $\sigma$ representing a spherical shell of collapsing matter and charge. The wall is described by only the radial degree of freedom, $R(t)$. The metric is taken to be the solution of Einstein equations for a spherical domain wall with charge. The metric is Reisner-Nordstr\"om outside the wall, as follows from spherical symmetry \cite{Lopez}
\begin{align}
ds^2= &-\left(1-\frac{2GM}{r}+\frac{Q^2}{r^2}\right) dt^2\nonumber\\
            & + \left(1-\frac{2GM}{r}+\frac{Q^2}{r^2}\right)^{-1} dr^2\nonumber\\
            & +r^2 d\Omega^2 \ , \ \ r > R(t)
\label{metricexterior}
\end{align}
where $M$ is the mass and $Q$ is the charge of the wall, respectively, and
\begin{equation}
d\Omega^2  = d\theta^2  + \sin^2\theta d\phi^2 \, .
\end{equation}
In the interior of the spherical domain wall, the line element is flat, as expected by Birkhoff's theorem,
\begin{equation}
ds^2= -dT^2 +  dr^2 + r^2 d\theta^2  + r^2 \sin^2\theta d\phi^2  \ ,
\ \ r < R(t)
\label{metricinterior}
\end{equation}
The equation of the wall is $r=R(t)$. The interior time coordinate, $T$, is related to the asymptotic observer time coordinate, $t$, via the proper time of an observer moving with the shell, $\tau$. The relations are
\begin{equation}
\frac{dT}{d\tau} =
      \left [ 1 + \left (\frac{dR}{d\tau} \right )^2 \right ]^{1/2}
\label{bigTandtau}
\end{equation}
and
\begin{equation}
\frac{dt}{d\tau} = \frac{1}{f} \sqrt{ f +
         \left ( \frac{dR}{d\tau} \right )^2 }
\label{littletandtau}
\end{equation}
where
\begin{equation}
f \equiv 1 - \frac{2GM}{R}+\frac{Q^2}{R^2}
\label{f}
\end{equation}

By integrating the Gauss-Codazzi equations of motion for the charged spherical domain wall derived in the previous section we find that the mass is a constant of motion (see also \cite{Lopez}) and is given by
\begin{equation}
 M= 4\pi \sigma R^2 \left[ \sqrt{1+R_\tau^2} - 2\pi G\sigma R\right] +\frac{Q^2}{2R}.
\label{Mass}
\end{equation}
The proof that $M$ is really a constant of motion is given in the Appendix. We therefore identify $M$ and the Hamiltonian of the system, i.e. $M\equiv H$.
The physical meaning of Eq. (\ref{Mass}) is straightforward. For a static shell, i.e. $R_\tau = 0$, the first term in square brackets is just the total rest mass of the shell. For a moving shell, $R_\tau \neq 0$ takes kinetic energy into account. The second term in
square brackets is the self-gravity or binding energy. Finally, the last term in (\ref{Mass}) is the electromagnetic contribution to the total mass (energy). In what follows, we will identify the conserved quantity (\ref{Mass}) with the Hamiltonian of the system.

It is also possible to take the non-relativistic large radius limit of the above hamiltonian.  For the case of constant mass $M_0=4\pi \sigma R^2$, the above hamiltonian becomes
\begin{equation}
H=M_0+ \frac{p^2}{2M_0}  -\frac{GM_0^2}{2R}+\frac{Q^2}{2R}
\end{equation}
which is the usual hamiltonian for a particle in a gravitational and electrical potential.  The extremal limit $M_0=\pm Q$ naturally corresponds to a free hamiltonian.  In this case it is clear that the identification of the conserved quantity with the hamiltonian is justified.

The collapse of the shell also obeys charge conservation, which is given by
\be
  D_aj^a=D_a(qu^q) ,
\ee
where $j^a$ is the four-current and $u^a$ is a timelike four-vector.

\section{Classical Equations of Motion}

In this section we will consider the classical equation of motion for the Reisner-Nordstr\"om domain wall (for earlier work see e.g. \cite{Campanelli:2003cs,Campanelli:2005xy,Gao:2008jy,Hawkins:1994sq}
and also \cite{Hajicek:1992pu,Hajicek:1998hd}).
To do so we consider an action that leads to the conserved hamiltonian. From Eq. (\ref{Mass}) the form of the action is then,
\begin{align}
  S_{eff}=-4\pi\int d\tau \sigma R^2\Big{[}&\sqrt{1+R_{\tau}^2}-R_{\tau}\sinh^{-1}(R_{\tau})\nonumber\\
      &-2\pi\sigma GR+\frac{Q^2}{8\pi\sigma R^3}\Big{]}
  \label{tau_action}
\end{align}
where $\tau$ is the propertime of the observer who is falling in with the shell and $R_{\tau}=dR/d\tau$. Now Eq. (\ref{tau_action}) can be written in terms of the asymptotic time $t$
\begin{align}
  S_{eff}=&-4\pi \int dt  \sigma R^2\Big{[}\sqrt{f-\frac{(1-f)}{f}R_t^2}\nonumber\\
  &-R_t\sqrt{f}\sinh^{-1}\left(R_t\sqrt{\frac{f}{f^2-R_t^2}}\right)\nonumber\\
  &+\left(\frac{Q^2}{8\pi\sigma R^3}-2\pi\sigma GR\right)\sqrt{f-\frac{R_t^2}{f}}\Big{]}
  \label{ExAction}
\end{align}
where $R_t=dR/dt$. From Eq. (\ref{ExAction}) the effective Lagrangian is then
\begin{align}
  L_{eff}=&-4\pi\sigma R^2\Big{[}\sqrt{f-\frac{(1-f)}{f}R_t^2}\nonumber\\
  &-R_t\sqrt{f}\sinh^{-1}\left(R_t\sqrt{\frac{f}{f^2-R_t^2}}\right)\nonumber\\
  &+\left(\frac{Q^2}{8\pi\sigma R^3}-2\pi\sigma GR\right)\sqrt{f-\frac{R_t^2}{f}}\Big{]}.
  \label{ExL}
\end{align}

The generalized momentum $\Pi$ can be derived from Eq. (\ref{ExL})
\begin{align}
  \Pi=&\frac{4\pi\sigma R^2}{\sqrt{f}}\Big{[}\frac{f^3R_t}{(f^2-R_t^2)\sqrt{f^2-(1-f)R_t^2}}\nonumber\\
  &+\frac{(1-f)R_t}{\sqrt{f^2-(1-f)R_t^2}}+\frac{(\frac{Q^2}{8\pi\sigma R^3}-2\pi\sigma GR)R_t}{\sqrt{f^2-R_t^2}}\nonumber\\
  &+f\sinh^{-1}\left(R_t\sqrt{\frac{f}{f^2-R_t^2}}\right)\Big{]}.
  \label{ExPi}
\end{align}
Thus the Hamiltonian in terms of $R_t$ is given by
\begin{align}
  H=&4\pi\sigma R^2\Big{[}\frac{f^3R_t^2}{(f^2-R_t^2)\sqrt{f^2-(1-f)R_t^2}}\nonumber\\
    &+f^2\left(\frac{(\frac{Q^2}{8\pi\sigma R^3}-2\pi\sigma GR)}{\sqrt{f}\sqrt{f^2-R_t^2}}+\frac{1}{\sqrt{f}\sqrt{f^2-(1-f)R_t^2}}\right)\Big{]}.
  \label{ExHam_t}
\end{align}

To obtain $H$ as a function of $(R,\Pi)$, we need to eliminate $R_t$ in favor of $\Pi$ using Eq. (\ref{ExPi}). This can, in principle, be done but is messy. Instead we consider the $R$ is close to $R_H$ and hence $f\rightarrow0$. In the limit $f\rightarrow0$ the denominators in Eq. (\ref{ExPi}) (and Eq. (\ref{ExHam_t})) are equal, therefore we can write
\be
  \Pi\approx\frac{4\pi\mu R^2R_t}{\sqrt{f}\sqrt{f^2-R_t^2}}
  \label{ExPi0}
\ee
where
\be
  \mu\equiv1+\frac{Q^2}{8\pi\sigma R_H^3}-2\pi\sigma GR_H
\ee
where $R_H$ is the horizon radius. Using Eq. (\ref{ExPi0}) we can then write the Hamiltonian as
\be
  H\approx \frac{4\pi\mu f^{3/2}R^2}{\sqrt{f^2-R_t^2}}\\
     =\sqrt{(f\Pi)^2+f(4\pi\mu R^2)^2}
     \label{ExH0}
\ee
and has the form of a relativistic particle, $\sqrt{p^2+m^2}$, with a position dependent mass term.

The Hamiltonian is a conserved quantity and so, from Eq. (\ref{ExH0}),
\be
  h=\frac{B^{3/2}R^2}{\sqrt{f^2-R_t^2}}
  \label{Exh0}
\ee
where $h=H/4\pi\mu$ is a constant. Solving Eq. (\ref{Exh0}) for $R_t$ we get
\be
  R_t=\pm f\sqrt{1-\frac{fR^4}{h^2}},
\ee
which, in the near horizon limit takes the form
\be
  R_t\approx\pm f\left(1-\frac{1}{2}\frac{fR^4}{h^2}\right)
\ee
since $f\rightarrow0$ as $R\rightarrow R_H$, where $R_H$ is the horizon radius.

The dynamics for $R\sim R_H$ can be obtained by solving the equation $R_t=\pm f$. Here we will consider two different cases, the non-extremal and extremal case.

\subsubsection{Non-Extremal Case}

For the non-extremal case, we consider the equality
\be
  1-\frac{2GM}{R_H}+\frac{Q^2}{R_H^2}=0
\ee
where $R_H$ is given by
\be
  R_H=GM\pm\sqrt{(GM)^2-Q^2}
  \label{R_H}
\ee
The plus sign is the outer and the minus sign is the inner horizon. To distinguish between them we write
\begin{align}
  R_+= & GM+\sqrt{(GM)^2-Q^2}\\
  R_-= & GM-\sqrt{(GM)^2-Q^2}
\end{align}
 Therefore we can then write Eq. (\ref{f}) as
\begin{align}
  f=&\left(1-\frac{GM+\sqrt{(GM)^2-Q^2}}{R}\right)\nonumber\\
      &\times\left(1-\frac{GM-\sqrt{(GM)^2-Q^2}}{R}\right)\\
      \equiv& f_{+}f_{-}.
\end{align}

Since in the near horizon limit $f\rightarrow0$, we can work with two different limits, either $f_{+}\rightarrow0$ or $f_{-}\rightarrow0$. In both cases $R_t\approx\pm f_{+}f_{-}$.For asymptotic observers watching the collapse, the horizon of interest is $f_{+}$. For $f_{+}\rightarrow0$, $f_{-}$ goes to a finite constant number, thus we have
\be
  R(t)\approx R_{+}+(R_0-R_{+})e^{\pm f_{-}t/R_{+}}
  \label{R_t_out}
\ee

We now make some comments on Eq. (\ref{R_t_out}). First, in the limit $(GM)^2>Q^2$, the exponential term in Eq. (\ref{R_t_out}) is positive definite. Thus it is easy to see that it takes the shell an infinite amount of time as seen by the asymptotic observer to reach $R_{+}$. In the limit of $(GM)^2>>Q^2$, $f$ reduces to $B$, where
\be
  B\equiv1-\frac{2GM}{R}.
\ee
This is the case studied in Refs. \cite{Vach_Stoj_Krauss} from the view point of an asymptotic observer and in Refs. \cite{Greenwood:2008zg,Greenwood:2008ht} from the view point of an infalling observer.

\subsubsection{Extremal Case}

In the extremal case $GM=Q$, so near the horizon we can write
\be
  f=\left(1-\frac{Q}{R}\right)^2\approx\left(\frac{R-Q}{R_h}\right)^2\sim\frac{\delta^2}{R_h^2} \, .
\ee
Here
\be
  R_h=GM=Q
\ee
is the position of the horizon in the extremal limit and
the small parameter $\delta$ is
\be
  \delta\equiv R-Q.
\ee
Therefore we can write
\be
  R_t\approx\delta_t=\pm\frac{\delta^2}{R_h^2}.
  \label{delta}
\ee
Solving Eq. (\ref{delta}) to leading order in $R-R_h$, the solution is
\be
  R(t)\approx R_h\pm(R_0-R_h)\frac{R_h^2}{t+R_h^2}.
  \label{R_delta}
\ee
Since we are interested in the collapsing case, we take the negative sign again. In that case, $R(t)=R_h$ only as $t\rightarrow\infty$.

From Eqs. (\ref{R_t_out}) and (\ref{R_delta}) we can see that again it takes an infinite amount of time for the shell to reach the horizon, as seen by the asymptotic observer.  However, in the extremal case the approach is not exponential, which is the consequence of the repulsive contribution of the charge.

\subsection{Cosmic censorship}

We now analyze what happens in the case when the parameters of the collapsing shell satisfy $Q^2>(GM)^2$. A black hole with such parameters is just a naked singularity, since the expression for the horizon
\be
r_h = GM \pm \sqrt{(GM)^2-Q^2}
\ee
would not be real. Thus, if the collapse proceeds all the way, this would represent a violation the cosmic censorship conjecture. However, this is not the case here. As we can see from Eq.~(\ref{R_t_out}), the solutions have a complex exponential, which implies oscillating solutions. The shell will collapse to some minimal radius at which the electromagnetic repulsion will overcome the gravitational attraction and cause the bounce. From that moment on, the shell will expand until it reaches some maximal radius at which the gravitational force will again dominate, and the new collapsing cycle will start. This is in agreement with the cosmic censorship conjecture.

\section{Quantum Effects Far from the Horizon}

Previously we have shown that the hamiltonian in the non-relativistic large radius limit is given by
\begin{equation}
H=M_0+ \frac{p^2}{2M_0}  -\frac{GM_0^2}{2R}+\frac{Q^2}{2R} \ .
\end{equation}
As this is similar to the usual Schrodinger equation for a hydrogen atom, bound state solutions are well known; bound states do not exist in the extremal limit.  Radial wavefunctions are given by Laguerre polynomials and the ground state energy is given by $E_0=-M_0 (GM_0^2-Q^2)^2/8$.  The factors of $2$ are due to the fact that this shell satisfies the Gauss-Codazzi equations.  The shell self interaction is not due to the full gravitational and electrical forces at the shell but the average of the values inside the shell and outside the shell.  The relativistic corrections to the energy as well as to the gravitational interactions can be further calculated perturbatively.

\section{Quantum Effects Near the Horizon}

\subsection{Near Horizon limit}

The classical Hamiltonian in Eq. (\ref{ExH0}) has a square root and so we first consider the squared Hamiltonian
\be
  H^2=f\Pi f\Pi+f(4\pi\mu R^2)^2
\ee
where we have made a choice for ordering $f$ and $\Pi$ in the first term. In general, we should add terms that depend on the commutator $[f,\Pi]$. However, in the limit $R\rightarrow R_H$, we find
\be
  [f,\Pi]\sim\frac{2}{R_H^2}\left(GM-\frac{Q^2}{R_H}\right) \ .
\ee
In the case of a charged black hole, the normal ordering ambiguity arises for black holes which are small relative to the Planck scale.  For uncharged black holes, the normal ordering ambiguity is less severe.  In fact in the extremal limit, the normal ordering ambiguity disappears completely and the hamiltonian is uniquely defined.

We now apply the standard quantization procedure. We substitute
\be
  \Pi=-i\frac{\partial}{\partial R}
\ee
in the squared Schr\"odinger equation,
\be
  H^2\Psi=-\frac{\partial^2\Psi}{\partial t^2}.
  \label{H2}
\ee
Then we have,
\be
  -f\frac{\partial}{\partial R}\left(f\frac{\partial\Psi}{\partial R}\right)+f(4\pi\mu R^2)^2\Psi=-\frac{\partial^2\Psi}{\partial t^2}.
\ee
To solve this equation, we define tortoise coordinates
\begin{align}
  u=&R+GM\ln\left|\frac{R^2}{2GMR-Q^2}-1\right|\nonumber
    & \\
&+\frac{2(GM)^2-Q^2}{\sqrt{(GM)^2-Q^2}} \ln (-\frac{R-2GM}{\sqrt{(GM)^2-Q^2}} )\nonumber &  \\
&-\frac{2(GM)^2-Q^2}{\sqrt{(GM)^2-Q^2}} \ln (2 -\frac{R-2GM}{\sqrt{(GM)^2-Q^2}} ) & \label{u}
\end{align}
%\begin{align}
%  u=&R+GM\ln\left|\frac{R^2}{2GMR-Q^2}-1\right|\nonumber\\
%    %&+\frac{2(GM)^2-Q^2}{\sqrt{Q^2-(GM)^2}}\tan^{-1}\left(\frac{R-GM}{\sqrt{Q^2-(GM)^2}}\right)
%    \label{u}
%\end{align}
which gives
\be
  f\Pi=-i\frac{\partial}{\partial u}.
\ee
Eq. (\ref{H2}) then gives
\be
  \frac{\partial^2\Psi}{\partial t^2}-\frac{\partial^2\Psi}{\partial u^2}+f(4\pi\mu R^2)^2\psi=0.
  \label{waveEq}
\ee
This is just the wave equation in a Minkowski background with a mass that depends on the position. From the structure of Eq. (\ref{u}), care needs to be taken to choose the correct branch since the region $R\in(R_H,\infty)$ maps onto $u\in(-\infty,\infty)$ and $R\in(0,R_H)$ onto $u\in(0,-\infty)$, where $R_H$ is given by Eq. (\ref{R_H}).

%We are interested in the situation of a collapsing wall. In the region $R\sim %R_H$, the logarithm in Eq. (\ref{u}) dominates and
%\be
%  R\sim e^{u/2GM}
%\ee
%since here the arctanh term goes to a constant. We look for wavepacket %solutions propagating toward $R_H$, that is, toward $u\rightarrow-\infty$. In %this limit
%\be
%  f\sim e^{u/GM}\rightarrow0
%\ee
%and the last term in Eq. (\ref{waveEq}) can be ignored. Wavepacket dynamics in %this region is simply given by the free wave equation and any function of %light-cone coordinates $(u\pm t)$ is a solution. In particular, we can write a %Gaussian wavepacket solution that is propagating toward the outer horizon
%\be
%  \Psi=\frac{1}{\sqrt{2\pi}s}e^{-(u+t)^2/2s^2}
%\ee
%where $s$ is some chosen width of the wave packet in the $u$ coordinate. The %width of the Gaussian wavepacket remains fixed in the $u$ coordinate while it %shrinks in the $R$ coordinate via the relation $dR=fdu$ which follows from Eq. %(\ref{u}).

We now turn to examine the quantization of incoming and outgoing states describing this infalling shell.  Fields propagating in this collapsing black hole background experience particle creation at a temperature given by the Hawking temperature.  There are many ways to derive this result including calculating the Bogolubov transformations between incoming and outgoing states at past and future null infinity.  One way to find the Bogolubov transformation is to consider the outgoing waves at future null infinity and using the high energy approximation, tracing these solutions back to past null infinity.  We now turn to this phenomenon and find that we can extract the necessary information directly from the horizon perturbations by considering a particular set of incoming and outgoing fields.

\subsection{Schwarzschild}

The tortoise coordinate $u$ for Schwarzshild is
\begin{equation}
u=R+ 2GM \ln |\frac{R}{2GM}-1| \ .
\end{equation}
In the near horizon limit, $R\approx 2GM$ we expand to lowest order
\begin{equation}
f=1-2GM/R \approx e^{u/2GM} \label{NHapprox}
\end{equation}
where $u\rightarrow -\infty$.
The equation of motion becomes after a further scaling of the coordinate $(t,u) \rightarrow 2GM (t,u)$
\begin{equation}
\frac{\partial^2 \psi}{\partial t^2} -\frac{\partial^2 \psi}{\partial u^2} + m^2 e^{u} \psi=0
\end{equation}
where the constant $m^2=(8 \pi \mu G M)^2$.  This wave equation is similar to the example of a time dependent mass examined in \cite{Strominger:2002pc, Maloney:2003ck}.  For sufficiently slowly varying exponent, the solution is essentially a plane wave and this occurs for large $GM$.  The general solution can be found by expanding $\psi$ in positive frequency plane waves
\begin{equation}
\psi(u,t)= R(u) e^{-i \omega t}
\end{equation}
in which case the Schrodinger equation is a form of the modified Bessel's equation
\begin{equation}
[\partial_u^2 - m^2 e^{u} + \omega^2] R(u)=0
\end{equation}
with two classes of normalizable solutions.  The modified Bessel functions
\begin{equation}
R_{in}^+(2 \sqrt{2} i m e^{u/2})=\frac{(im)^{i\omega}}{\sqrt{2 \omega}} \Gamma(1-2i\omega) J_{-2i \omega} (2\sqrt{2} i m e^{u/2})
\end{equation}
in the near horizon limit $u\rightarrow -\infty$ become the positive frequency solutions
\begin{equation}
R_{in}^+\approx \frac{e^{-i\omega u}}{\sqrt{2\omega}} \ .
\end{equation}
Altogether the effective potential vanishes near the horizon and the wave equation becomes that of a free massive field with the standard ingoing (and outgoing) Fourier modes.  At distances far from the horizon these correspond to solutions which grow without bound and independently of $\omega$
\begin{equation}
R_{out}^+\approx e^{-u/4+2\sqrt{2} m e^{u/2}}
\end{equation}
although this exponential growth in the modes is an artifact of our approximation Eq.~\ref{NHapprox}, which no longer is valid far from the horizon.  In fact in the large radius limit, the solutions become wavelike.

In addition we consider the solutions
\begin{equation}
R_{out}^+=\sqrt{\frac{\pi}{2}} (i e^{2\pi \omega})^{-1/2} H^2_{-2i\omega} (2\sqrt{2} i\sqrt{m} e^{u/2})
\end{equation}
which are related by a Bogolubov transformation
\begin{equation}
R_{out}^+=a R_{in} + b \bar{R}_{in} \ .
\end{equation}
to the incoming modified Bessel functions
\begin{equation}
H_{-2i\omega}^2 (x)= \frac{J_{2i\omega}(x) - e^{2\pi \omega} J_{-2i\omega}(x)}{-i \sin(-2i\pi \omega)} \ .
\end{equation}
Near the horizon $u\rightarrow - \infty$, the Hankel function for complex argument can be expanded as
\begin{equation}
H_{-2i\omega}^2 (2\sqrt{2} i \sqrt{m} e^{u/2})\approx e^{-i\omega u}- e^{2\pi \omega} e^{i \omega u} \ .
\end{equation}
Relating the incoming and outgoing states we find $R_{out}^+ \approx (R_{in}+e^{-2\pi \omega} \bar{R}_{in})$.  This physically shows particle production as the shell approaches the horizon and that the density of particles created is given by the ratio
\begin{equation}
|b/a|^2=e^{-4\pi \omega}
\end{equation}
at temperature $1/4\pi$.  Restoring the units of temperature we find that the modes of the shell experience a temperature
\begin{equation}
T=1/8\pi GM
\end{equation}
which is the Bekenstein-Hawking temperature of the black hole!  The Bogolubov transformation also shows that at large $\omega$ the vacuum for the in and out states are identical.

The two sets of wavefunction solutions are related by the Bogolubov transformation. Despite the fact that the outgoing state is a pure state, it has a Boltzmann distribution at temperature $T_H=1/8\pi G M$ which is precisely the Hawking temperature of the black hole.  This is a result of the fact that the interaction potential is periodic in Euclidean space.

Normally the Hawking temperature is measured at asymptotic infinity while the near horizon temperature is blue-shifted and infinite.  In our near horizon analysis we extract the finite Hawking temperature for outgoing modes although it is measured relative to that of the incoming modes.  Naturally this should be a finite quantity.  If one were to take the usual temperature calculation and simultaneously measure the outgoing radiation versus the incoming radiation at large but finite radial distance then the relative blueshifting would cancel out leading to the same Hawking temperature.

We note here that this result implies outgoing thermal radiation, though we did not consider any matter fields propagating in the background of the collapsing shell. The shell itself is both the source of gravitational field and matter that collapses.  Ref.~\cite{Kraus:1994by} performed an analysis of a massless spherical shell and the radiation it emits.  In their analysis the shell was an approximation of a particle moving in the black hole background.  Thus, we may conclude that the shell itself loses its mass, perhaps in the form of emitted gravitons and pairs of spherical shells.

\subsection{Charged Reissner-Nordstrom}

For the charged black hole the analysis is very similar except that two logarithm terms are kept in the expansion near the outer horizon and the distance from the outer horizon is written in terms of the coordinate $u$ as
\begin{equation}
R-R_+\approx e^{u/(GM + \frac{2(GM)^2-Q^2}{2 \sqrt{(GM)^2-Q^2}}) }\ .
\end{equation}
Up to rescalings the equations of motion are the same modified Bessel equations.  Performing the same analysis in terms of the near horizon modes, the temperature experienced by the gravitational modes is the same as the Bekenstein-Hawking temperature for a charged black hole
\begin{equation}
T=\frac{\sqrt{(GM)^2-Q^2}}{2 \pi (GM+\sqrt{(GM)^2-Q^2})^2} \ .
\end{equation}

\subsection{Scalar Fields and Temperature}

In the above analysis we have argued that the Hawking radiation temperature can be calculated for the quantum mechanical wavefunction describing gravity and the collapsing shell near the horizon.  In this picture it was not necessary to invoke the asymptotically flat region of the black hole.  The relevant thermal properties could be calculated in the vicinity of the horizon.  Furthermore no blue shift factors were needed either as both the incoming and outgoing states were localized to the horizon.

The same analysis can also be applied to a scalar field theory in the vicinity of a black hole.  For simplicity we will work with the Schwarzschild black hole. A scalar field $\Phi$ can be decomposed into modes of the form $\phi=r^{-1} f(r,t) Y_{lm}(\theta, \phi)$ so that the wave equation becomes
\begin{equation}
\frac{\partial^2 f}{\partial t^2} -\frac{\partial^2 f}{\partial u^2} +(1-\frac{2GM}{r})[\frac{l(l+1)}{r^2}+\frac{2GM}{r^3}+m^2] f=0
\end{equation}
where we have  $u$ is the tortoise coordinate.  In the near horizon limit $(1-2M/r) \approx e^{u/2GM}$ as before and the second term in parenthesis becomes a constant.

The mode equation of the scalar field therefore is just the modified Bessel equation that we found for the shell.  It is then possible to use the same analysis to find the incoming and outgoing states which can be written as modified Bessel functions.  These outgoing states are thermal relative to the incoming states and are at the Hawking temperature.

% is the following function and its complex conjugate
%\begin{equation}
%\psi_{\vec{p}} = \frac{2^{-i \omega}}{\sqrt{2 \omega}} \Gamma (1-i\omega) e^{i %\vec{p}\cdot \vec{x}} J_{-i \omega} (e^t)
%\end{equation}
%In the limit where $t\rightarrow -\infty$,
%\begin{equation}
%\psi \rightarrow \frac{1}{\sqrt{2\omega}} e^{-i\omega t + i \vec{p} \cdot %\vec{x}}
%\end{equation}

\section{Quantum Effects Near the Singularity}
\subsection{Near singularity limit for the uncharged black hole}

In this section we investigate the question of quantum effects when the collapsing shell approaches the origin (i.e. classical singularity at $R\rightarrow 0$). An observer using time $t$ can not study this limit, so we perform our analysis using the time $\tau$ of an observer located on the collapsing shell. The Hamiltonian (in terms of $R_{\tau}$) is just the conserved quantity in (\ref{Mass}).  After setting $Q=0$, the effective Lagrangian consistent with the conserved quantity (\ref{Mass}) is
\begin{equation}
  L=-4\pi\sigma R^2\left[\sqrt{1+R_{\tau}^2}-R_{\tau}\sinh^{-1}(R_{\tau})-2\pi\sigma GR\right].
  \label{Lagrangian}
\end{equation}
The generalized momentum, $\Pi$, can be derived from this Lagrangian as
\begin{equation}
  \Pi=4\pi\sigma R^2\sinh^{-1}(R_{\tau}).
  \label{momentum}
\end{equation}
The Hamiltonian in terms of $R_{\tau}$ is
\begin{equation}
  H=4\pi\sigma R^2\left[\sqrt{1+R_{\tau}^2}-2\pi\sigma GR\right]
\end{equation}
From the Hamiltonian we can get $R_\tau$ as
\be \label{rtau}
R_\tau = \pm \sqrt{\left( \frac{h}{R^2} + 2\pi \sigma GR \right)^2-1}
\ee	
where $h=H/{4\pi \sigma}$.
Here we can study two cases. First we consider the ultra-relativistic limit near the origin where $R_\tau $ is very large.
%Then, near the origin, i.e. in the limit of $R\rightarrow 0$ the %classical solution for $R_\tau $  (keeping only the leading order %term) becomes
%\be \label{ras}
%  R_{\tau} \approx-\frac{h}{R^2} .
%\ee
Up to the leading term near the origin, the Hamiltonian is
\begin{equation}
  H=4\pi\sigma R^2 R_{\tau} .
\end{equation}
Clearly, this choice eliminates the Newton's constant G from the equation, and thus important gravitational effects are not included.
In terms of the generalized momentum (\ref{momentum}), the Hamiltonian now is
\begin{equation}
H=4\pi \sigma R^2 \sinh (\frac{\Pi}{4 \pi \sigma R^2}) \ .
\end{equation}
The Schrodinger equation becomes
\begin{equation}
2\pi \sigma R^2 [e^{(\frac{\Pi}{4 \pi \sigma R^2})} -e^{-(\frac{\Pi}{4 \pi \sigma R^2})}] \psi(R,\tau) = i \frac{\partial \psi(R,\tau)}{\partial \tau}
\end{equation}
Defining a new variable $u=R^3$, the equation becomes
\begin{equation}
2\pi \sigma u^{2/3} [e^{-(\frac{3i}{4\pi \sigma} \frac{\partial}{\partial u})} - e^{(\frac{3i}{4\pi \sigma} \frac{\partial}{\partial u})}] \psi (u,\tau) =i \frac{\partial \psi(u,\tau)}{\partial \tau}
\end{equation}
Since the exponentials are now just the translation operators, we have
\begin{equation}
2\pi \sigma u^{2/3} [\psi (u - \frac{3i}{4\pi \sigma},\tau) - \psi (u + \frac{3i}{4\pi \sigma},\tau)]=i \frac{\partial \psi}{\partial \tau} \ .
\end{equation}
It is interesting to note that this equation is dependent on $\sigma$ which describes the particular shell.  This in general might be solvable from a recursion relationship.
This is a manifestly non-local equation. The non-locality that we found may be a simple consequence of the fact that we are using the functional Schrodinger formalism which is only an effective theory, i.e. only an approximation of some more fundamental local theory. The other possibility is that the quantum description of the black hole physics requires inherently non-local physics. The answer to this question requires further investigation.

Leaving this question aside, we examine a particular local limit as follows.  When $\sigma \rightarrow \infty$ the above expression becomes a derivative
\begin{equation}
2\pi \sigma (-3i/2\pi \sigma)  u^{2/3} \partial_u \psi (u, \tau) = i\frac{\partial \psi}{\partial \tau}
\end{equation}
or after simplifying
\begin{equation}
-3 u^{2/3} \partial _u \psi =\partial_\tau \psi .
\end{equation}
Although the left side of this equation appears to vanish in the near singularity limit, by rewriting this equation in terms of the variable $R$ we find the simple relationship
\begin{equation}
-\partial _R \psi = \partial _\tau \psi
\end{equation}
which has solutions of the form $\psi= \psi(R-\tau)$. In order to maintain the same position on the profile, if $R$ becomes smaller $\tau$ has to move to smaller values as well, which means that the wavefunction moves backwards in time.  However inside the black hole horizon, time and space switch their roles.  By analyzing the conformal structure of the black hole, we find that inside the horizon $\tau$ moving to smaller values corresponds to a normal infalling trajectory.  Our local equation for the evolution of the wavefunction appears simple to understand in this limit.  In this case though there is no boundary condition imposed on the wavefunction and it appears that the wave equation allows solutions to propagate through the singularity.  This is due to the fact that this limit corresponds to the pure kinetic energy limit which neglects gravity.

We now consider the case when the tension $\sigma$ is very large and the gravitational interaction term in Eq. (\ref{rtau}) is dominant
\be
2\pi \sigma GR  \gg \frac{H/4\pi \sigma}{R^2}  \geq 1 \ .
\ee	
In this case $R_\tau \approx 2\pi \sigma G R$ and the hamiltonian becomes $H\approx 4\pi \sigma R^2/R_\tau$.  The above inequality can be written as $ 8\pi \sigma^2 G R^3 \gg H $ and substituting for the hamiltonian we find that this is the ultra-relativistic limit
\be
(\sigma G R)^2 \gg 1 \ \  \leftrightarrow  \ \ R_{\tau}>>1 \ .
\ee
In terms of the generalized momentum (\ref{momentum}), the Hamiltonian now is
\begin{equation}
H=\frac{4\pi \sigma R^2}{ \sinh (\frac{\Pi}{4 \pi \sigma R^2})} \ .
\end{equation}
which we write using the translation operator
\be
\frac{8\pi \sigma u^{2/3}}{T(-3i/4\pi\sigma)-T(3i/4\pi\sigma)}\psi (u,\tau)=i\frac{\partial \psi(u,\tau)}{\partial \tau}
\ee
Taking the large $\sigma$ limit, reduces the denominator to $-6i/4\pi\sigma\partial_u$ so the equation of motion becomes an inverse differential operator equation
\be \label{ido}
\frac{16\pi^2 \sigma^2 u^{2/3}}{3 \partial_u} \psi (u,\tau)=\frac{\partial \psi(u,\tau)}{\partial \tau} .
\ee
Here, the wavefunction is non-infinite and constant at the origin provided that the inverse differential operator has a finite behavior.
This indicates that quantum effects may be able to remove the classical gravitational singularity at the center. To make a definite statement one would need to calculate the conserved probability for the whole space (not only in the near horizon and near the classical singularity limit), which is proportional to $\psi^* \psi$ but also contains a non-trivial measure term due to the curved background. The integrated probability over the whole space-time should then be non-singular. Within our approximations we can not do this (we do not have solutions which are valid everywhere), but the fact that the wave function is not singular at $R=0$, where the classical singularity was located, is a strong indication that quantum effects may make gravity non-singular.

Eq. (\ref{ido}) also indicates that strong gravity regime is manifestly non-local (because of the inverse differential operator). While in the previous ultra-relativistic limit with the absence of G we were able to remove non-local effects by taking large $\sigma$ limit, in this case we are not able to do the same.
This indicates that gravity is inherently non-local, and while quantum mechanical non-localities may be removed by taking an infinitely large measuring apparatus, once we turn on gravity this is no longer possible \footnote{This is along the lines presented by N. Arkani Hamed at  "The Workshop on Tests of Gravity and Gravitational Physics, May 19-21, 2009, Case Western Reserve University, Cleveland, Ohio.}.

\subsection{Near singularity limit for the charged black hole}

When we include charge, the full Hamiltonian from (\ref{Mass}) is
\begin{equation}
H=4\pi \sigma R^2 [ \sqrt{1+R_\tau^2} -2\pi G \sigma R] + \frac{Q^2}{2R} \ .
\end{equation}
From here we find
\begin{equation}
R_\tau= \sqrt{(\frac{H}{4\pi \sigma R^2}-\frac{Q^2}{8\pi\sigma R^3} + 2\pi G \sigma R)^2-1}
\end{equation}
which is in the limit of small radius dominated by the charge, i.e. $R_\tau\approx Q^2/8\pi \sigma R^3$.  In this limit it is possible to drop the gravitational energy term in the Hamiltonian and get the relativistic kinetic energy limit
\begin{equation}
H = 4\pi \sigma R^2 R_\tau+\frac{Q^2}{R} \ .
\end{equation}
The generalized momentum in this limit becomes
\begin{equation}
\Pi =4\pi \sigma R^2 \sinh^{-1} R_\tau
\end{equation}
so the Hamiltonian can be written as
\begin{equation}
H= 4\pi \sigma R^2 \sinh (\frac{\Pi}{4\sigma R^2}) +\frac{Q^2}{R} \ .
\end{equation}
The Schrodinger equation is the same as before but with the addition of the Coulomb term
\begin{equation}
2\pi u^{2/3} [\psi (u - \frac{3i}{4\pi \sigma},\tau) - \psi (u + \frac{3i}{4\pi \sigma},\tau)]+ \frac{Q^2}{2 u^{2/3}}\psi=i \frac{\partial \psi}{\partial \tau}
\end{equation}
and we can study this in the limit of large energy density $\sigma \rightarrow \infty$ to get
\begin{equation}
-\partial _R \psi -i\frac{Q^2}{2R}\psi= \partial _t \psi
\end{equation}
which has solutions of the form
\begin{equation}
\psi=\psi_0 R^\frac{-i Q^2}{2} e^{A(R-\tau)}
\end{equation}
where $A$ is an arbitrary constant. There is a phase factor $R^{-iQ^2/2}$ which corresponds to an infinite number of modulus one oscillations.  The exponential term corresponds to the generic wave which propagates to the classical singularity at $R=0$.

\section{Conclusion}

In this paper we have analyzed the collapse of a massive shell of all perfect fluids with charge using the Gauss-Codazzi method.  The hamiltonians and lagrangians were constructed for these systems and several limits were analyzed.  The key point was to then invoke the functional Schrodinger formalism to find key quantum features.  The near horizon limit analysis led to a new way to determine the Hawking temperature of black holes without reference to asymptotic states.  While the first order terms led to thermal radiation it may be useful to further study higher order terms to look for non-thermality. The near singularity limit showed that the behavior of the wavefunction was non-singular.
While the equations to solve were inherently non-local and contained an infinite number of derivatives, certain local limits (i.e. very massive shells) were analyzed.

Questions regarding the wavefunction remain.  Is there a measure for the wavefunction which leads to the conservation of probability?  In particular how is the probability conserved on spatial slices when timelike/spacelike notions change through the horizon.  It is also unclear if the probability of the wavefunction inside the shell is finite in the limit where the shell collapses to zero size.  If it is, how do we treat this information which is within the shell.  Can fluctuations outside the shell, propagate and stay inside the shell?

It would be interesting to see if one can find numerical solutions to the quantum wavefunction in all regions of spacetime.

We note that we worked in the context of the functional Schrodinger formalism which in many ways resembles the Wheeler-DeWitt approach to quantum gravity \cite{DeWitt:1967yk}. Related work in the existing literature based on different approaches also indicates that quantum effects may be capable of removing classical singularity at the center  \cite{Bogojevic:1998ma,Modesto:2004xx,Guendelman:2009vz,Bronnikov:2006fu,Shankaranarayanan:2003qm,MankocBorstnik:2005ib,Yeom:2008qw}. While it has been previously argued that 
the standard Schrodinger formalism does not yield conclusive claims about the non-singularity \cite{Husain:1987am}, our work apparently gives strong indications for non-singular behavior at the center of the black hole.

\appendix
\section{}

Here we prove that the mass given by Eq.~(\ref{Mass}) is the constant of motion.
From Eqs.~(14) and (15) one finds that the acceleration for the charged domain wall is given by
\be
  \frac{1}{\alpha}R_{\tau\tau}=\frac{Q^2}{8\pi\sigma R^4}-\frac{2\alpha}{R}+6\pi\sigma G
\ee
where $\alpha$ is defined as
\be
  \alpha=\sqrt{1+R_{\tau}^2}.
\ee
From Eq.~(22) we have that the mass of the domain wall is given by
\be
  M=4\pi\sigma R^2\left(\alpha-2\pi\sigma GR\right)+\frac{Q^2}{2R}.
\ee
To show that the mass is a conserved quantity it is sufficient to show that $M_{\tau}=0$. So using Eq.~(22) we have
\begin{align*}
  M_{\tau}=&R_{\tau}\Big{[}-\frac{Q^2}{2R^2}+8\pi\sigma R\left(\alpha-2\pi\sigma GR\right)\nonumber\\
          &+4\pi\sigma R^2\left(\frac{1}{\alpha}R_{\tau\tau}-2\pi\sigma G\right)\Big{]}
\end{align*}
Now using the acceleration we can write this as
\begin{align*}
  M_{\tau}=&R_{\tau}\Big{[}-\frac{Q^2}{2R^2}+8\pi\sigma R\left(\alpha-2\pi\sigma GR\right)\nonumber\\
          &+4\pi\sigma R^2\left(\frac{Q^2}{8\pi\sigma R^4}-\frac{2\alpha}{R}+6\pi\sigma G-2\pi\sigma G\right)\Big{]}
\end{align*}
Multiplying out and canceling terms leaves
\be
  M_{\tau}=R_{\tau}(0)=0,
\ee
hence the mass is a conserved quantity.

\end{document}